\documentclass[12pt]{article}
\usepackage{amsmath,amssymb,amsfonts}
\usepackage{graphicx}
\usepackage{booktabs,colortbl}
\usepackage{cancel}
\usepackage{subfigure}

\newcommand{\be}{\begin{equation}} 
\newcommand{\ee}{\end{equation}} 
\newcommand{\bea}{\begin{eqnarray}}  
\newcommand{\eea}{\end{eqnarray}}
\newcommand{\bs}{\begin{split}} 
\newcommand{\es}{\end{split}}

\newcommand{\units}[1]{~\mathrm{#1}}

\newcommand{\TeV}{~\mathrm{TeV}}

\newcommand{\ctoprule}{\toprule[0.5mm]}
\newcommand{\cbottomrule}{\bottomrule[0.5mm]}

\newcommand{\cmrule}{\midrule[0.25mm]}

\newcommand{\MET}{$E_T{\hspace{-0.47cm}/}\hspace{0.35cm}$}

\begin{document}



\vspace*{3mm}

\begin{center}

\renewcommand{\thefootnote}{\fnsymbol{footnote}}
\setcounter{footnote}{1}

{\Large {\bf The Least Supersymmetric Signals at the LHC
}} \\
\vspace*{0.75cm}
{\bf J.\ de Blas}\footnote{E-mail: jdeblasm@nd.edu},
{\bf A.\ Delgado}\footnote{E-mail: antonio.delgado@nd.edu}
and
{\bf B.\ Ostdiek}\footnote{E-mail: bostdiek@nd.edu}

\vspace{0.5cm}

Department of Physics, University of Notre Dame,\\
Notre Dame, IN 46556, USA

\end{center}
\vspace{.5cm}

\begin{abstract}
 
\noindent We study the implications at the LHC for the minimal (least) version of the supersymmetric standard model. In this model, supersymmetry is broken by gravity and extra gauge interaction effects, providing a spectrum similar in several aspects to that in natural supersymmetric scenarios. Having the
first two generations of sparticles partially decoupled means that any significant signal can only involve gauginos and the third family of sfermions. In practice, the signals are dominated by gluino production with subsequent decays into the stop sector. As we show, for gluino masses below 2300 GeV, a discovery at the LHC is possible at $\sqrt{s}=14$ TeV, but will require large integrated luminosities.
\end{abstract}

\renewcommand{\thefootnote}{\arabic{footnote}}
\setcounter{footnote}{0}


\section{Introduction}
\label{section_Intro}

The high sensitivity of the electroweak scale to large energy scales has been one of the main
motivations to postulate the existence of physics beyond the Standard Model (SM), with supersymmetry (SUSY) one of the most appealing candidates. Moreover, attending to naturalness arguments, such new physics should not be far from the TeV scale. Apart from being an elegant solution to the hierarchy problem, the minimal supersymmetric standard model (MSSM) provides a candidate for dark matter, and has other nice features such as gauge coupling perturbative unification. On the other hand,  new problems arise when models with SUSY have to face existing theoretical and experimental constraints. Some of these are quite generic, while others depend significantly on the mechanism of supersymmetry breaking. 

Fine-tuning and the little hierarchy problem belong to the class of endemic problems. Indeed, the current value of the Higgs mass is somewhat larger than the tree-level MSSM prediction, and thus we have to rely on large radiative corrections from the third generation of sfermions to raise $m_h$ to the observed value. In this regard, the first and second generations of sparticles do not play an important role. This means that at least one of the stops needs to be around $1-2$ TeV, while all the other sfermions can be heavier ($\sim 10$ TeV) without introducing much more fine tuning. Thus, from the low energy point of view, it can be argued that only the third generation needs to be supersymmetric.
Notice also that current searches at the large hadron collider (LHC) have been cornering minimal setups involving SUSY, imposing quite stringent bounds on the first generations of squark masses \cite{ATLAS:2012sma}. Hence, considering a heavier first and second generation is actually well motivated from the phenomenological point of view. 

Another problem in supersymmetric theories with general soft-breaking parameters is the flavour problem. Flavour data imposes very strong constraints on the structure of the soft parameters for the first two families. In particular, this problem affects some specific SUSY breaking mechanisms, like gravity mediation \cite{Nilles:1983ge}. Other scenarios, like gauge mediation \cite{Giudice:1998bp}, are free from this flavour problem, since the breaking is only communicated through gauge interactions. This, however, suffers from another hierarchy problem (which can be naturally solved within gravity mediated SUSY breaking \cite{Giudice:1988yz}), the so-called $\mu-b_\mu$ problem. In general, no single SUSY breaking mechanism is completely satisfactory by itself, but some of their flaws can be cured by combining several different sources of SUSY breaking \cite{Combine}.

In Ref. \cite{Delgado:2011kr} a simple minimal model addressing all the previously discussed issues was proposed. This combines all the good features of both gravity mediation and a gauge mediation-like mechanism, absent of flavour problems. It also preserves gauge unification. The (extra) gauge interactions only affect the first two families, whose sparticles get large degenerate masses $\sim 10$ TeV. Thus, at the TeV scale, the effective theory is only supersymmetric in the gauge, Higgs and third family sectors. Hence, this is named the least supersymmetric standard model (LSSM). The characteristic features of the spectrum, with the absence of the first and second generations in the low energy phenomenology, share resemblance with other supersymmetric constructions, like natural/effective/more minimal supersymmetry \cite{NatEffMoreMinSUSY}. In this paper we present a more phenomenological analysis of the model in Ref. \cite{Delgado:2011kr}. We study in more detail the phenomenology at the LHC, and in particular we focus on strongly produced signals. 

In the next section we briefly review the main theoretical aspects of the LSSM. We also
comment on the viability of the model in terms of reproducing electroweak symmetry breaking 
consistently with the current experimental bounds, and discuss the resulting spectra. The main signals and the phenomenology at the LHC are studied in Section \ref{section_LHC}. Finally, we present our conclusions.


\section{The Least Supersymmetric Standard Model}
\label{section_LSSM}

We will work within the simple scenario discussed in Ref. \cite{Delgado:2011kr}, and we refer to that reference for more details. The model consists of a supersymmetric version of a $U(1)^\prime$
extension of the SM. The extra charges, $Q^\prime$, are given in Table \ref{tab:charges}. In that table,
apart from the MSSM fields we have also included a gauge singlet $S$, and  two extra chiral fields, $\varphi_{1,2}$, whose scalar components acquire a vacuum expectation value (vev) $\left<\varphi_{1,2}\right>=\hat{v}$, breaking the  $U(1)^\prime$ symmetry. The resulting construction is anomaly free, as anomalies cancel between the first and second generations.
\begin{table}[!t]
\begin{center}
\begin{tabular}{c c c c c c c c }
\toprule
\multicolumn{1}{c}{}&$\psi_1$&$\psi_2$&$\psi_3$&$H_{u,d}$&$\varphi_1$&$\varphi_2$&$S$\\
\midrule
$Q^\prime$&$+1$&$-1$&$0$&$0$&$+1$&$-1$&$0$\\
\bottomrule
\end{tabular}
\caption{$U(1)^\prime$ charges used in the model. $\psi_i=(q_i,l_i,u^c_i,d^c_i,e^c_i)$, $i=1,2,3$, denote the three SM generations.}
\label{tab:charges}
\end{center}
\end{table}

There are two sources of SUSY breaking in the model. The first is gravity mediation, which is universal. Secondly, there is another (secluded) sector where SUSY is broken at a scale $M_*$ by a chiral field $X=M_*+\theta^2 F$, with $\sqrt{F}\ll M_*$. This extra source of SUSY breaking is communicated to the visible sector through the $U(1)^\prime$ interactions. Thus, only the first and second SM generations are sensitive to these effects at the leading order, acquiring soft masses
\be
\hat{m}^2=\frac{\hat{g}^2(M_*)}{128 \pi^4}\frac{F^2}{M_*^2},
\ee
with $\hat{g}$ the $U(1)^\prime$ gauge coupling constant. This is also the same order as the $U(1)^\prime$ gaugino mass, $M_{\hat{\lambda}}\sim \hat{m}$. After $U(1)^\prime$ symmetry breaking, the vector multiplet $(\hat{A}_\mu,\mbox{Re}(\varphi_1-\varphi_2),\hat{\lambda},\mbox{Re}(\tilde{\varphi}_1-\tilde{\varphi}_2))$ and the chiral multiplet $(S,\varphi_1+\varphi_2,\tilde{S},\tilde{\varphi}_1+\tilde{\varphi}_2,\tilde{S})$ get masses of ${\cal O}(\hat{v})$. These also receive small corrections from SUSY breaking, of ${\cal O}(\hat{m})$. Finally, the gravitino mass is given by $m_{3/2}\simeq k F/\sqrt{3}M_P$, where $M_P$ is the Planck scale and we will consider the theory-dependent numerical prefactor $k\sim {\cal O}(1)$. Since gravity is the only interaction communicating SUSY breaking to the gauge and Higgs sectors, as well as the third generation, all the soft parameters are of order $m_{3/2}$. Moreover, $\mu \simeq m_{3/2}$ can also be easily explained via the Giudice-Masiero mechanism \cite{Giudice:1988yz}. Thus, $m_{3/2}$ has to be of electroweak size, but large enough to generate a third generation of squarks in the TeV region so we can explain a Higgs mass around $125$-$126$ GeV \cite{Aad:2012tfa,Chatrchyan:2012ufa}.

Notice that apart from providing the same satisfactory explanation to the $\mu$-$b_\mu$ problem as in gravity mediation, this combined scenario has also naturally suppressed flavour changing neutral currents (FCNC). Indeed, the approximate degeneracy between the first two families and the relative large mass splitting with the (lighter) third generation helps in suppressing FCNC operators.

\subsection{Electroweak symmetry breaking and the Higgs mass}

This particular implementation of the LSSM is completely specified by eight parameters (and the sign of $\mu$). First we have the scales $m_0$, $M_{1/2}$ and $A_0$, that fix the gravity-mediation contribution to the soft scalar masses, gaugino masses and $a$ terms, respectively, at the ultraviolet scale. We will choose this to be the grand unification scale $M_{\mbox{\tiny GUT}}$, defined by $g_1(M_{\mbox{\tiny GUT}})=g_2(M_{\mbox{\tiny GUT}})$. The $U(1)^\prime$-mediation SUSY breaking parameters include $F/M_*$ and $M_*$. The other $U(1)^\prime$ parameters are the gauge coupling constant $\hat{g}$ and the symmetry breaking vev $\hat{v}$. Some of these parameters can be bounded or related by different arguments \cite{Delgado:2011kr}. First, we will trade the SUSY breaking scale $F/M_*$ for the common soft mass scale $\hat{m}$, which can be bounded by fine-tuning arguments to be $\hat{m}\lesssim 10 \units{TeV}$ for any value of $M_*\lesssim M_{\mbox{\tiny GUT}}$. Demanding $\hat{m}\simeq 10\units{TeV}$ as well as $m_{3/2}$ such that the third generation of sfermions is around $1\units{TeV}$ we can roughly estimate the size of the scale at which SUSY is broken and communicated by the $U(1)^\prime$ interactions, in terms of the grand unification scale: 
$$M_*\simeq \hat{g}^2 M_{\mbox{\tiny GUT}}/4\pi\simeq 10^{15}\units{GeV}.$$
In the last equality, we have assumed for the gauge coupling $\hat{\alpha}=\hat{g}^2/4\pi \simeq 1/20$. Assuming perturbative values for the yukawa interactions, the ratio $\hat{v}/M_*$ can be bounded by flavour constraints, $\hat{v}/M_*\lesssim 10^{-2}$ . We will take $\hat{v}/M_*=10^{-2}$. Finally, with the value of $\tan{\beta}$, all the freedom of the model is fixed.

In order to study electroweak symmetry breaking for this particular implementation of the LSSM, we make use of a two-loop renormalization group analysis \cite{Martin:1993zk} to run the parameters from their values at the ultraviolet down to low energies, where we use the existence of the electroweak symmetry breaking vacuum to determine the remaining MSSM parameters ($|\mu|$ and $b_\mu$) and compute the physical spectrum. In the running we consider several different scales where part of the spectrum is decoupled. From the grand unification scale (where gravity mediation boundary conditions are set) down to $M_*$ we consider the full MSSM spectrum plus the $U(1)^\prime$ vector multiplet, the extra chiral fields $\varphi_{1,2}$, $S$, and the messenger fields mediating SUSY breaking through the $U(1)^\prime$ interactions. The latter occurs at the scale $M_*$ where we decouple the messenger fields, and add the $U(1)^\prime$ gauge-mediation contributions to the soft parameters, in particular to the first and second generations of sfermion masses. Below $\hat{v}$ the extra gauge boson, gaugino and the chiral fields $\varphi_{1,2}$ and $S$ are decoupled, leaving only the MSSM field content. Finally, below $\hat{m}$, the first two families of sfermions are decoupled. The resulting theory is evolved down to low energies, where we minimize the effective scalar potential, including the one-loop and leading two-loop radiative corrections. In order to minimize the leading ${\cal O}(y_t^4)$ corrections, this last step is performed at a matching scale $M_{\tiny \mbox{match}}=\sqrt{m_{\tilde{t}_1}m_{\tilde{t}_2}}$. The existence of the electroweak vacuum, as well as the computation of the particle spectrum, have been cross-checked using SuSpect 2.4 \cite{Djouadi:2002ze}.

We have checked that, although it always requires some degree of fine-tuning (inevitable as in many minimal supersymmetric extensions of the SM), a 125-126 GeV Higgs can be found consistently with spectra where the stops and gluinos masses are as light as the current experimental constraints: $m_{\tilde{t}_1}\gtrsim 600-700 \units{GeV}$ \cite{StopAtlas} and  $M_{\tilde{g}}\gtrsim 1200-1300\units{GeV}$ \cite{GluinoCMS,GluinoAtlas}. Notice, however, that current limits on stop masses \cite{StopAtlas} only cover the region for $m_{\chi^0_1}\lesssim 200-250$ GeV. In our model, on the other hand, the lightest neutralino is typically mostly bino and its mass is around those values or heavier. Indeed,  the constraints on the gluino mass push the parameter $M_{1/2}$ towards relatively large values, yielding large neutralino masses. Points with stops even lighter than 600 GeV and heavy $\chi_1^0$, which are allowed in our model, are still consistent with the experimental results. 

As is typically expected in this kind of models, the lightest stop is always mostly right handed. This can be understood from the hierarchy between the beta functions for right-handed and left-handed stop soft masses. The corresponding hierarchy also explains why $\tilde{b}_R$ is heavier than $\tilde{b}_L$.\footnote{Indeed, at the leading order, for moderate values of $\tan{\beta}$ and $M_{1/2}\lesssim m_0,~\!A_0$ we have, $\beta_{m^2_{u_3^c}}>\beta_{m^2_{q_3}}>\beta_{m^2_{d_3^c}}$. Notice though that, because of the large masses for the first and second generations, their leading two-loop effects can have some impact in this one-loop relation.} Since in order to reproduce the adequate Higgs mass at least one of the stops must be significantly heavy, $m_{\tilde{t_2}}\gtrsim 1-2 \units{TeV}$, sbottom masses are bounded to be quite large. This implies that, even though there exist regions where the lightest sbottom mass is below $M_{\tilde{g}}$, for gluino masses accessible at the LHC ($M_{\tilde{g}}\lesssim 2\units{TeV}$), gluino decays will be in general dominated by top/stop final states.

\begin{figure}[t]
\begin{center}
\includegraphics[width=0.9\textwidth]{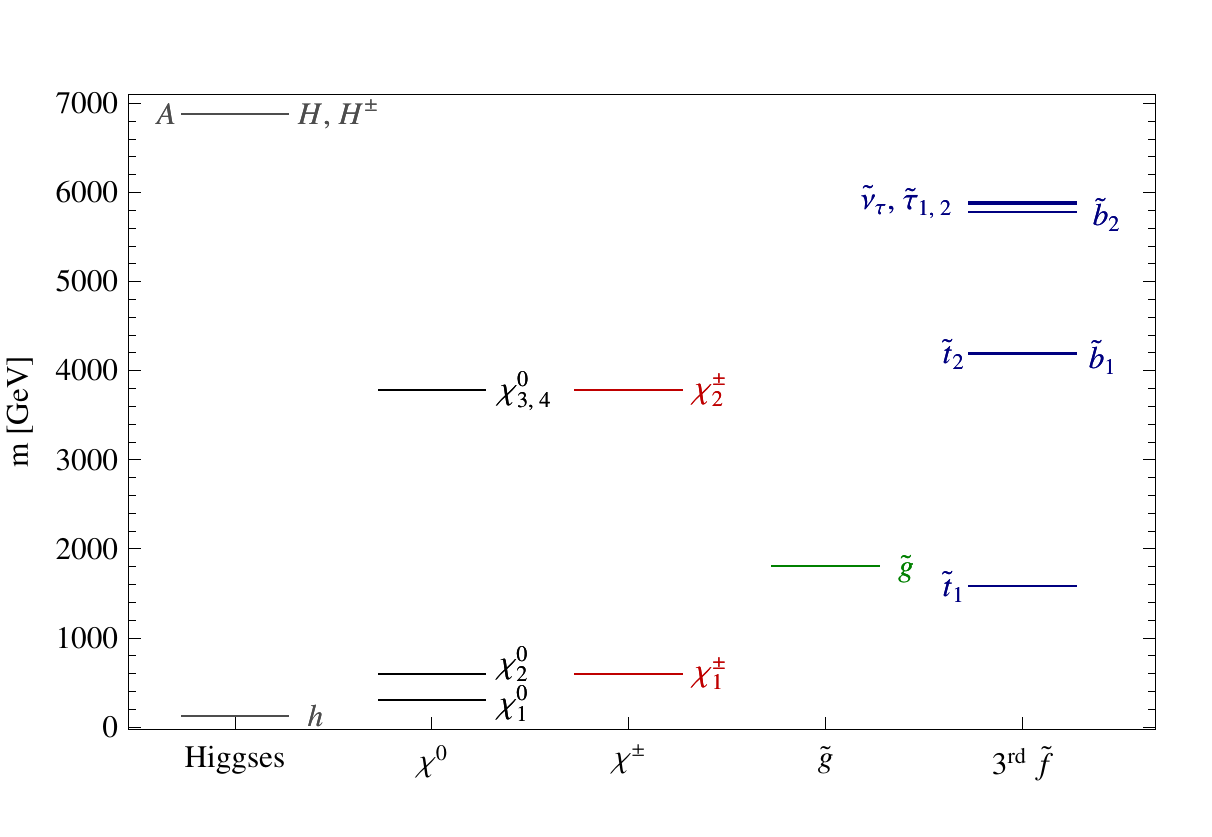}
\caption{Higgs and supersymmetric particle spectrum for the benchmark point A introduced in this section.}
\label{fig_Benchpt_A}
\end{center}
\end{figure}
In order to illustrate the phenomenological features of this model we choose a generic benchmark point with 
\be
m_0=6 \units{TeV},~~~M_{1/2}=0.65\units{TeV},~~~A_0=-10.2\units{TeV},
\ee
a positive sign for $\mu$ ($\mu=3.8\TeV$) and $\tan{\beta}= 10$. As explained above, the $U(1)^\prime$-mediation scale is set to $\hat{m}=10\units{TeV}$. We will refer to this as point A. The resulting spectrum contains a lightest CP even Higgs boson of 125.4 GeV, consistent with the experimental limits. The values for all the non-SM particles are illustrated in Figure \ref{fig_Benchpt_A}. Because of the large value of $\mu$, the lightest neutralino/chargino states are gaugino-like. In particular, the lightest supersymmetric particle (LSP) is the lightest neutralino (mostly Bino).  As explained in the next section, the most characteristic (and efficient) signal to look for at the LHC involves gaugino production and decays into the stop sector. In this point, the lightest stop has a mass of around 1600 GeV, while the left-handed states, which are much heavier, are around 4200 GeV. The gluino is relatively heavier than the lightest stop, with a mass of $\sim 1800$ GeV.

For latter convenience, let us also introduce a very similar point, referred to as point B from now on. The values of the input parameters are
\be
m_0=5.8 \units{TeV},~~~M_{1/2}=0.85\units{TeV},~~~A_0=-9.85\units{TeV},
\ee
and the same sign for $\mu$ and value for $\tan{\beta}$. These parameters have been purposely chosen so that the spectrum is essentially the same as point A, but with somewhat heavier gauginos. In particular, we set $M_{1/2}$ so the gluino mass is around $ 2250$ GeV.

\section{LHC signals}
\label{section_LHC}

Given the particular features of this model, the list of observable signals at the LHC is quite short. As explained in Ref. \cite{Delgado:2011kr}, the decoupling of the first and second generations of sfermions reduces the possible decays of charginos and neutralinos to
$$\chi^\prime\rightarrow \left\{\begin{array}{l}\chi~W/Z\\ \chi~h\\ f\tilde{f}~(f=\tau, t, b)\end{array}\right..$$
Thus, leptonic signals can only come from the decays of the $W$ or $Z$ and the multijet$+\cancel{E}_T$ signal is much enhanced compared to standard MSSM scenarios. Still, neutralino/chargino production, being of electroweak size, is not the most efficient way of testing signals for this model ($\sigma(pp\rightarrow \chi^{0(\pm)}+\mbox{X})=0.7~\! (2.5) \units{ab}$). Indeed, the leading signal is gluino pair production,  with subsequent decays into the third generation states
$$pp\rightarrow \tilde{g}\tilde{g},~~~
\tilde{g}\rightarrow \left\{\begin{array}{l}t \tilde{t}_1\rightarrow b \bar{b}~W^+ W^-~\chi^0_1\\
b \tilde{b}_1\rightarrow b\bar{b}~\chi_1^0\end{array}\right.,$$
where in the stop decay chain we have used the fact that, since the lightest stop is mostly right handed, charged decay modes $\tilde{t}_1\rightarrow b \chi_1^\pm$ are highly suppressed, and the decay is dominated by the channel $\tilde{t}_1\rightarrow t \chi_1^0$, see Table \ref{tab_st1Width}. Moreover, for the benchmark points presented in the last section, gluino decays into bottom/sbottom are not allowed, since the lightest sbottom is much heavier than $\tilde{g}$. As explained in the previous section, even if sbottom masses below the gluino mass are possible in the allowed region of the parameter space, gluino decays are still expected to be dominated by top/stop final states. Therefore, the channel discussed here will offer the clearest signal.

We have also considered, for completeness, stop pair production. For the points we are discussing, since the lightest stop has a fairly large mass, the resulting $pp\rightarrow \tilde{t}_1\tilde{t}^{~\!\!*}_1$ cross section seems to be too small to consider this an efficient search channel. For instance, for the LHC at $\sqrt{s}=14\units{TeV}$, we find for the point A, prior to any cuts, $\sigma(pp\rightarrow \tilde{t}_1\tilde{t}^{~\!\!*}_1)= 0.1\units{fb}$ while $\sigma(pp\rightarrow \tilde{g}\tilde{g})= 1.612\units{fb}$. These numbers, as well as all the new physics signals presented in this section have been computed using Madgraph 5 \cite{Alwall:2011uj}, with the model implemented via Feynrules \cite{Christensen:2008py,Duhr:2011se}. While the previous numbers correspond only to the partonic cross sections, the results presented below use Pythia 6 \cite{Sjostrand:2006za} for the hadronization and PGS \cite{pgs} to simulate detector effects. Finally, for the computation of some of the SM backgrounds, we also use Alpgen \cite{Mangano:2002ea}. Regarding stop pair production, points with lower stop masses, and therefore larger cross sections, are allowed by current constraints. As stressed in the previous section, points in the parameter space where the stop masses can be as low as $\sim 600-700$ GeV can be found, while gluino masses are constrained by experimental limits to be above 1200-1300 GeV. In this case the production cross section for stop pairs can be indeed larger than that of gluino pairs. However, for light stops, even if the production cross sections are fairly large, the kinematic distributions are similar to those of the leading SM background, $t\bar{t}$ production, making it very difficult to clearly distinguish the signal. For not so-light stops, like the ones in our model, the distributions can be somewhat different, but the cross sections decrease rapidly as we increase $m_{\tilde{t}_1}$. Moreover, as explained above, the experimental limits on $M_{\tilde{g}}$ indirectly constrain the lightest neutralino mass to relatively large values, $m_{\chi^0_1}\gtrsim 200$ GeV, in the region of the $m_{\chi^0_1}-m_{\tilde{t}_1}$ plane where experimental searches are less sensitive to the new physics signal. Because of these issues, we expect that the phenomenology of the model would manifest first in gluino pair production. In what follows we will focus our attention on this channel.

\begin{table}[t]
\centering
ÊÊÊÊ\begin{tabular}{c | c c }
	\ctoprule
ÊÊÊÊÊÊÊ	Decay		&	Width	&	Branching\\ 
ÊÊÊÊÊÊÊ	channel		&	[GeV]         	&	Ratio \\ 
	\cmrule
	$t~\!\chi^0_1$	&	3.191	&	0.995	\\
	$t~\!\chi^0_2$      &	0.005	&	0.002	\\
	$b~\!\chi^\pm_1$&	0.009	&	0.003	\\
	\cmrule
	Total			&	3.205	&		\\
	\cbottomrule
	\end{tabular}
\caption{Widths of the decay modes of the lightest stop, $\tilde{t}_{1}$, for $m_{\tilde{t}_{1}}\approx 1600 \units{GeV}$.\label{tab_st1Width}}
\end{table}

\begin{figure}[ht]
\centering
\subfigure[]{
\includegraphics[width=12cm]{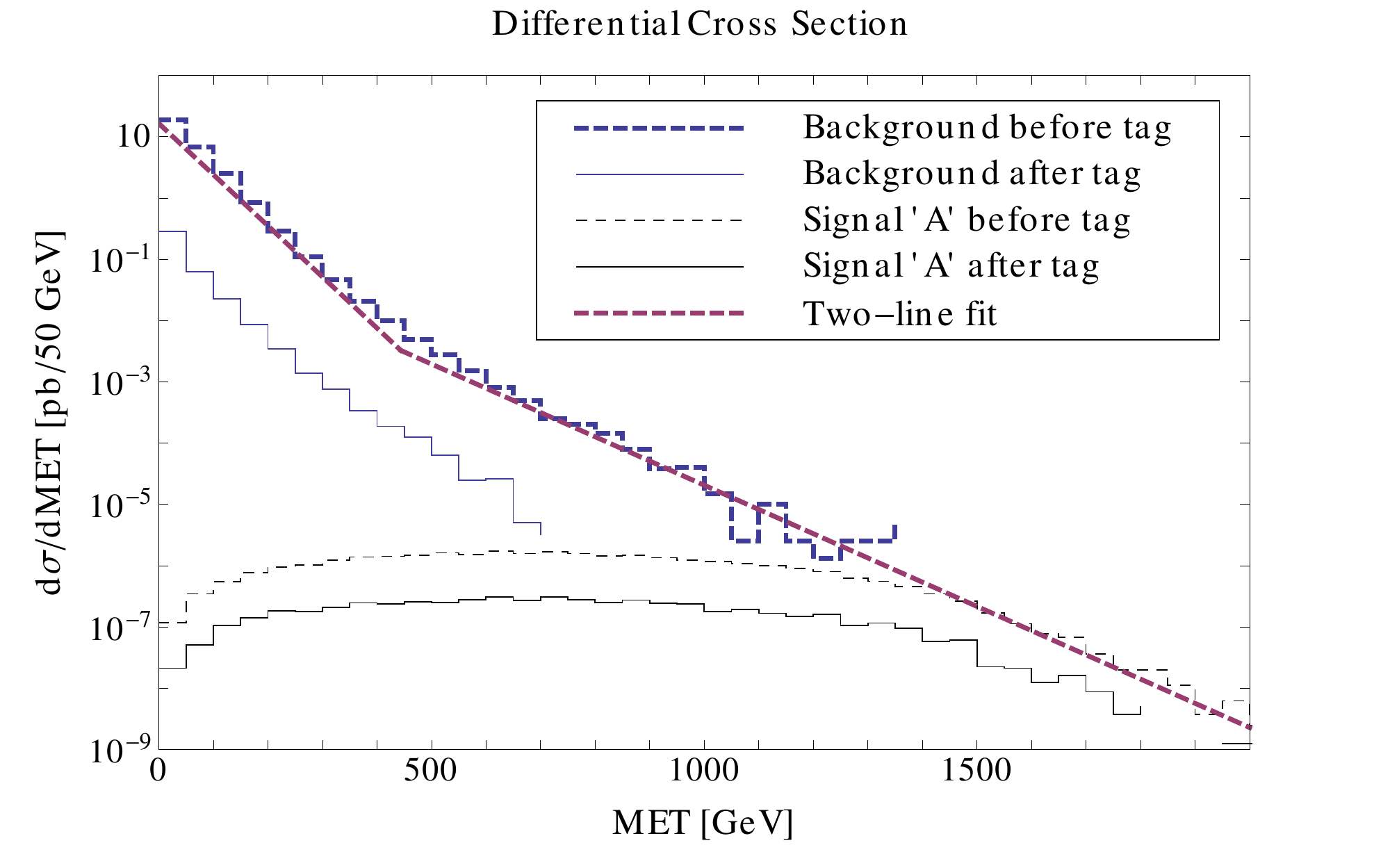}
\label{fig_dcsa}
}
\subfigure[]{
\includegraphics[width=12cm]{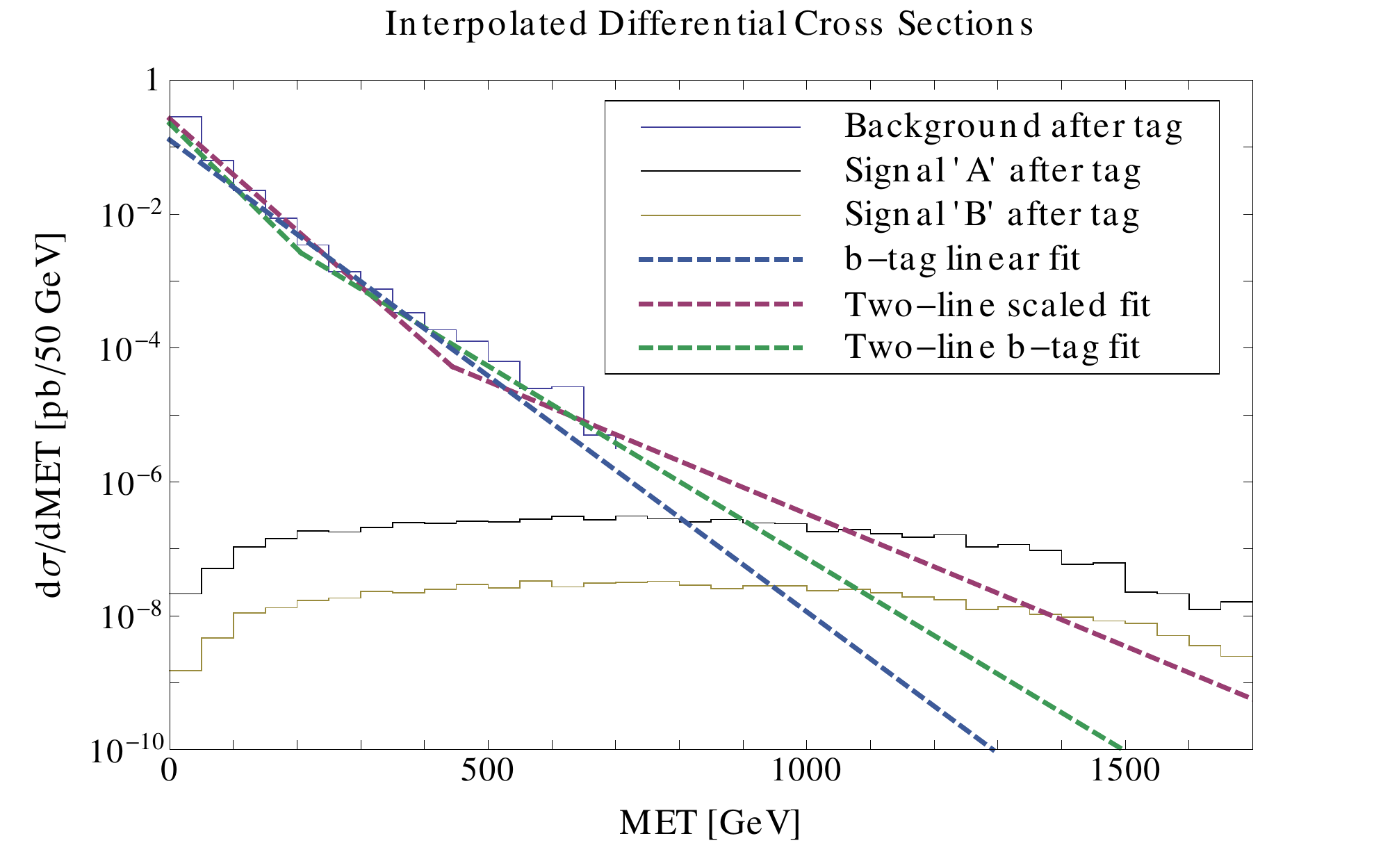}
\label{fig_dcsb}
}
\caption{Differential cross section as a function of the \MET for $\sqrt{s} = 14$ TeV. (a) Comparison between signal and background before and after (tight) $b$ tagging. (b) Signal and background after $b$ tagging, together with different estimations for the background in the region \MET$\gtrsim 600$ GeV.\label{Fig_dcs}}
\end{figure}

The signal we are looking for has four top quarks plus missing energy. At the detector level, this turns into four $b$-jets + four other jets/leptons + missing energy. The SM backgrounds for the signal are $t\overline{t}$ + jets, $t\overline{t}+W/Z$ + jets, $W/Z$+jets, di-boson+jets, and $q\overline{q}q\overline{q}$ where $q = (t,b)$. The $W/Z$+jets, di-boson+jets and $q\overline{q}q\overline{q}$ backgrounds are negligible because they either do not produce energetic enough jets to be mistagged as a $b$-jet or do not contain enough missing energy. As such, these are not included in the following. The backgrounds considered are $t\overline{t}+W/Z$ + jets (calculated with Madgraph) and the dominant $t\overline{t}$ + jets (calculated with Alpgen). The total cross section for the background at $\sqrt{s}=14$ TeV is 1477 pb, while the cross section for the signal in point A (B) is 1.612 fb (0.170 fb). However, as the signal leads to four $b$-jets at the detector level, we can improve the signal/background discrimination by using $b$ tagging. Because of the $b$-tag efficiency, this of course will also reduce the signal. Moreover, due to detector limitations, we do not expect to be able to tag each $b$ jet. Thus, we require at least three $b$ tags, and use a loose $b$-tag performance\footnote{One can use different $b$-tag performances: {\it loose}, with a higher $b$-tag efficiency but also higher probability for a light jet miss-tagging, and {\it tight}, with lower efficiency but also lower fake rate. The exact numbers for these efficiencies can be found in the PGS documentation \cite{pgs}. While the use of a tight $b$-tag yields a larger reduction of the background, this effect does not outweight the penalization on our signal, and this becomes more evident as we look for heavier gluino masses.}. We also demand to see at least four other jets and no photons in the final state. Table \ref{tab_xsec1} provides the results for both signal and background before and after the three $b$ tags and jet/photon restrictions.

\begin{table}[t]
\begin{center}
\begin{tabular}{l c c }
\ctoprule
		& Before $b~\!$-tag	& After $b~\!$-tag \\
\cmrule
	Signal Point A	&	1.612 fb	& 0.286 fb	\\
	Signal Point B	&	0.170 fb	& 0.032 fb	\\
	Background	&	1477 pb & 19.18 pb\\
\cbottomrule
\end{tabular}
\caption{Signal and background cross sections before and after applying $b$ tagging. Points A and B differ in the gluino mass: $M_{\tilde{g}}\approx 1.8, 2.25$ TeV, respectively.\label{tab_xsec1}}
\end{center}
\end{table}

As usual, the presence of the LSP at the end of the decay chains translates into a large amount of missing energy. We have plotted the differential cross section as a function of the missing transverse energy (MET or \MET) in Figure \ref{Fig_dcs}. As can be observed, the distribution for the signal is characterized by being somewhat flat, extending up to around 1800 GeV. In simulating the background, several technical issues make it difficult to generate the corresponding distribution up to such large energies, especially after requiring the three $b$ tags. After the background is run through Pythia and PGS, it extends up to around 600 GeV (1500 GeV before $b$-tagging). To deal with this, we use different estimation methods, explained below, and we will always assume a conservative point of view when analyzing the results. First, Figure \ref{fig_dcsa} illustrates the above-mentioned necessity of $b$ tagging, as otherwise the background typically dominates over all the range of \MET. In that figure we also show the results for signal and background after $b$ tagging, for comparison, as well as an estimation for the original background based on two linear fits to data\footnote{This is a five parameter fit, where not only the slopes and intercepts of both lines are determined, but also the point where both lines cross each other.}. Figure \ref{fig_dcsb} focuses on the results after the $b$ tag. In this case, in order to give an estimation for the background we have used several different methods. First, we use a simple linear fit. Secondly, we also perform an analogous fit to two lines, as in the case before $b$ tagging. Finally, we have scaled down the background estimation before $b$ tagging to fit the $b$ tagged data in the region where both are available. At any rate, we observe that a discrimination between signal and background should be possible, but demanding enough significance for a discovery may require a large integrated luminosity. 

The detailed results using the different estimation methods, as well as different $b$-tagging performances, are summarized in Table \ref{tab_counts}. In particular, we show in that table the number of events and the significance obtained for an integrated luminosity of 200 fb$^{-1}$ for point A. Being conservative, we can thus claim that an observation of this signal at the LHC at $\sqrt{s}=14\units{TeV}$ would require collecting luminosities $\gtrsim 200\units{fb}^{-1}$. Although this is quite a large amount of data, it is still well within the LHC luminosity projections by the end of its operation. Point B marks the LHC reach for this kind of search. As can be seen from the results in Table \ref{tab_counts}, even for an integrated luminosity of 1 ab$^{-1}$ the discovery of gluinos with masses $\sim 2300$ GeV would be challenging.
\begin{table}[t]
\begin{center}
\begin{tabular}{l c c c c c c}
\ctoprule
	Estimation & $E_T^{\tiny\mbox{Cut}}{\hspace{-0.68cm}/}\hspace{0.35cm}$&~~$\sigma_{\tiny \mbox{B}}^{\tiny \mbox{Estimated}}$& $\sigma_{\tiny \mbox{S}}$	& S	& B	& S$/\sqrt{\mbox{B}}$ \\
	Method & [GeV] & [ab]& [ab]	&\multicolumn{3}{c}{${\cal L}=200\units{fb}^{-1}~(1000\units{fb}^{-1})$} \\
\cmrule
	Linear	             &\phantom{0}850 \phantom{0}(950)	         & 17.1 (3.73)   & 106.6 (10.8)	         & 21 (11) & 3 (4)	& 11.5 (5.6) \\
&&&&&&\\[-0.25cm]
	Two-Line &\phantom{0}950 (1100)	& 10.4 (1.43)   &\phantom{0}80.7 (7.01)	&16 \phantom{0}(7)    & 2 (1)	& 11.2 (5.9)\\
&&&&&&\\[-0.25cm]
	Two-Line &1100 (1400)      & 14.7 (0.96)   &\phantom{0}50.3 (2.26)	& 10 \phantom{0}(2)	      & 3 (1)	& \phantom{0}5.9 (2.3) \\
(Scaled)&&&&&&\\
\cbottomrule
\end{tabular}
\caption{Results after applying three $b$-tags for the point A. Results for point B are given in parentheses. The `linear' estimation method fits a line to the $b$-tagged data. The `Two-line' finds the best fit of two lines to the $b$-tagged data. The `Two-line (Scaled down)' method scales down the Two-line fit to the background before $b$ tagging to fit the $b$-tagged data. The energy at which the estimated background crosses the signal is given by $E_T^{\tiny\mbox{Cut}}{\hspace{-0.68cm}/}\hspace{0.35cm}$. Finally, the rounded number of events and significance for point A (B) are for 200 fb$^{-1}$ (1000 fb$^{-1}$) of data.\label{tab_counts}}
\end{center}
\end{table}%


\section{Conclusions}
\vspace{-0.155cm}

Having in mind the results from current searches at the LHC, in Ref. \cite{Delgado:2011kr} a simple scenario containing the minimal set of parameters in the MSSM that are consistent with theoretical and phenomenological constraints was introduced. The spectrum of the model is characterized by gravity mediation-like masses for the Higgs, gaugino and third family sectors, while the first and second generations are pushed up to $\sim10 \units{TeV}$ by SUSY breaking contributions mediated by extra gauge interactions. In this short paper we have studied the main signals of this model that can manifest at the LHC. These are characterized by the absence of the first and second generations in the low energy phenomenology. We have focused our analysis on strongly produced signals, as electroweak processes offer less chance for a clear discovery. In particular, we focus on
gluino pair production, which is expected to be much more clear than the production of stop pairs. 
\vspace{-0.05cm}

We have studied different benchmark points, and here we have presented some representative results. We choose one point where both gluinos and stops are heavy, $\gtrsim 1.5\units{TeV}$, in order to illustrate the LHC reach for this model when the leading signal is gluino pair production. In this case we observe that, being conservative, gluinos $\sim 1800\units{GeV}$ would be observable at $\sqrt{s}=14 \units{TeV}$, provided we have large luminosities $\gtrsim 200 \units{fb}^{-1}$. Also from our results, we can infer that the LHC would not be sensitive to gluinos heavier than $\sim 2300$ GeV.
\vspace{-0.05cm}

Let us finally remark that, as opposed to other standard MSSM-like models where the first families of squarks and sleptons offer a more rich phenomenology, there are no other places where this model could clearly manifest at the LHC. Thus, an excess in the discussed channels together with the absence of any other signals might be a hint that this kind of scenarios is being realized in nature. On the other hand, it would not be an easy task to distinguish from other similar constructions where only the third family is relatively light.

\section*{Acknowledgements}

This work has been supported in part by the U.S. National Science Foundation under Grant PHY-1215979. This research was also supported by the Notre Dame Center for Research Computing through computational resources.



\end{document}